\documentclass[final]{aipproc}
\layoutstyle{6x9}

\begin{document}

\title
     [Magnetic Effects Change Our View of the Heliosheath]
     {Magnetic Effects Change Our View of the Heliosheath}

\author{M. Opher}{
    address={Jet Propulsion Laboratory, MS 169-506, 
4800 Oak Grove Drive, Pasadena, CA 91109},
    email={merav.opher@jpl.nasa.gov}
}
   
\iftrue
\author{P. C. Liewer}{
   address={Jet Propulsion Laboratory, MS 169-506, 
4800 Oak Grove Drive, Pasadena, CA 91109},
    email={merav.opher@jpl.nasa.gov},
}

\author{M. Velli}{
   address={Dept. of Scienza del Spatizo Firenzi, Italy},
}

\author{T. I. Gombosi}{
   address={Space Physics Research Laboratory, Department of Atmospheric, 
Ann Arbor, MI},
}

\author{W.Manchester}{
   address={Space Physics Research Laboratory, Department of Atmospheric, 
Ann Arbor, MI},
}

\author{D. L. DeZeeuw}{
    address={Space Physics Research Laboratory, Department of Atmospheric, 
Ann Arbor, MI},
}

\author{G. Toth}{
      address={Space Physics Research Laboratory, Department of Atmospheric, 
Ann Arbor, MI},
}

\author{I. Sokolov}{
     address={Space Physics Research Laboratory, Department of Atmospheric, 
Ann Arbor, MI},
}
\fi

\begin{abstract}
There is currently a controversy as to whether Voyager 1 has already crossed the 
Termination Shock, the 
first boundary of the Heliosphere. 
%An important aspect of 
%this controversy is our poor understanding of this region. 
The region between the Termination Shock and 
the Heliopause, the Helisheath, is one of the most unknown regions theoretically. 
In the Heliosheath magnetic effects are crucial, as the solar magnetic field is compressed at the 
Termination Shock by the slowing flow. 
Recently, our simulations showed that the 
Heliosheath presents remarkable dynamics, with turbulent flows and the presence of a jet flow at the 
current sheet that is 
unstable due to magnetohydrodynamic instabilities \cite{opher,opher1}. In this paper we review 
these recent results, 
and present an additional simulation with constant neutral atom background. In this case the jet is still present 
but with reduced intensity. 
Further study, e.g., including neutrals and the tilt of the solar rotation from the magnetic
axis, is required before we can definitively address how the Heliosheath behaves. Already we can say that this region
presents remarkable dynamics, with turbulent flows, indicating that the Heliosheath might be very different from what we previously 
thought.
\end{abstract}
`
\date{\today}

\maketitle
 
\section{Introduction}
As the Sun travels relative to the interstellar medium with a velocity of approximately 25km/s\cite{frisch},
it is subject to an interstellar wind.
The basic structures that are formed by the interaction between the solar 
wind and the supersonic interstellar wind are: the Termination Shock (TS), the Heliopause (HP), and, 
possibly a Bow Shock (BS). Figure 1, taken from one of our simulations, is a 3D view of the 
global heliosphere showing the Parker spiral (white lines) being pulled tailward. 
The red contours denotes the BS and the yellow the TS. The black lines 
follow the flow streamlines and the HP is the location where they start to bend.
The region between the TS and the HP, the Heliosheath, is one of the most mysterious and unknown regions. 
In this region, the magnetic field is crucial as the solar magnetic field is compressed at the TS by the slowing flow. 
The solar magnetic field reverses polarity at the 
heliospheric current sheet (HCS). One of the major questions is how the HCS behaves beyond the TS.
Recent observations indicate that Voyager 1, now at 90AU (reached in Nov 05, 2003), is in a region unlike 
any encountered in its 26 years of exploration  \cite{krimi,mcdonald,burlaga}. 
There is currently a controversy as to whether Voyager 1 has already crossed the TS. 
An important aspect of this controversy is our poor understanding of this region.
What do we want to know about the structure of the Heliosheath? We whould like to tackle the 
following fundamental questions among others: 1) The type of 
flows; 2) The fate of the current sheet beyond the TS; 
3) The role of the magnetic field; 4) The turbulence level;
and 5) The distribution of ionized and neutral particles.

In previous hydrodynamic models, the region beyond the TS has a constant plasma 
pressure and temperature and the heliospheric boundary was a smooth, rounded surface. This is not 
true if the solar magnetic field is included: the plasma pressure, temperature and density 
downstream the shock are not uniform and constant, and the heliospheric boundary is highly 
distorted from the rounded appearance of the hydrodynamic models. Figure 2 shows side by side the 
standard view (with no solar nor interstellar magnetic field) and a case from a recent 
simulation \cite{opher1} where the solar magnetic field is included. The HP bulges out and 
the BS is pushed farther out. 
%\begin{figure*}[ht!]
%\begin{minipage}[t] {0.5\linewidth}
%\begin{center}
%\includegraphics[angle=0,scale=0.3]{fig2a.ps}
%\end{center}
%\end{minipage} \hfill
%\begin{minipage}[t] {0.5\linewidth}
%\begin{center}
%\includegraphics[angle=0,scale=0.26]{fig2b.ps}
%\end{center}
%\end{minipage} \hfill
%\caption{Contours of pressure for (a) Case with no magnetic field (b) With magnetic field with 
%resolution of $0.75AU$ at the current sheet.}
%\label{fig2}
%\end{figure*}
There have been several numerical approaches to tackle this complicated interaction. Much work has 
focused on the careful treatment of the neutral component, using a kinetic treatment, without including magnetic 
field effects. There are very few works that included both the solar and interstellar magnetic fields 
in a self consistent way in three-dimensional geometry. The drawback of these models is that 
the neutrals are treated with a fluid approximation. In short, currently there {\it is no} model yet able
to include the major ingredients and describe properly the global heliosphere, 
especially the Heliosheath.
In our recent studies \cite{opher,opher1}, we performed a 3D MHD modeling using the 
adaptive grid BATSRUS code, developed by the Univ. of Michigan, 
including the solar magnetic field with an unprecedented grid resolution. Our model did not include effects such as
the tilt of the magnetic to the solar rotation axis. However, we already obtained 
new phenomena, e.g., the formation of an unstable jet flow at the current sheet, beyond the TS. 
These results show how magnetic effects are crucial and can change 
the view that we have of the Heliosheath.
The outline of the paper is the following: In the first and second sections 
we summarize our recent results \cite{opher,opher1}. The third section present results 
from out most recent simulation, where we included a constant neutral background. 
Finally the forth section presents a discussion on the present status of our knowledge and future work.

\section{Formation of a Jet at the Current Sheet}
Nerney, Suess \& Schmahl \cite{nerney,nerney1}, in analytic studies of the region beyond the TS, 
predicted the presence of magnetic ridges due to the compression of the azimuthal interplanetary field. 
Their studies were made in the kinematic approximation where the magnetic field back reaction on the 
flow was neglected. 
Our recent results in 3D MHD simulations \cite{opher,opher1} confirm the presence of 
the magnetic ridges beyond the TS (seen also in \cite{linde}). 
Besides the formation of the magnetic ridges, we found that 
a jet-sheet 
forms. In the current-sheet region, due to the absence of an azimuthal magnetic component, there is no magnetic pressure 
to slow down the flow and the solar wind streams with a higher velocity. 
This leads to the formation of a jet in the meridional plane and 
sheet in the equatorial plane. Due to the shear between the flow in the ``jet-sheet'' and the 
flow in the surrounding medium, the jet-sheet (the current sheet) becomes unstable. We were the first to 
report this phenomena \cite{opher}. In the jet region, the wind velocity is much faster than 
the surrounding medium. At $x=-210 AU$, for example, the flow streams at the equator with 
a velocity $\sim 150km/s$, while the surrounding medium flows with a velocity of $40km/s$. The jet pushes 
solar material aside that starts to flow toward the TS. This produces turbulent vortices (see Figure 3b). 
We used an adaptive mesh refinement 
allowing us to get to spatial resolutions previously not obtained (on the order of $1.5AU$ and $0.75AU$, respectively, in Opher et al. 
\cite{opher,opher1}) at the HCS. Using such high 
resolution and extending the refined region, we were able to resolve the jet extending to 
150 AU beyond the TS.

Why didn't the previous studies \cite{linde,washimi} see it? 
Figure 3 shows two cases. One with spatial resolution of $3AU$, and the other 
with resolution of $0.75~AU$ at the current sheet. 
\begin{figure}[!b]
 \resizebox{.4\columnwidth}{!}
  {\includegraphics{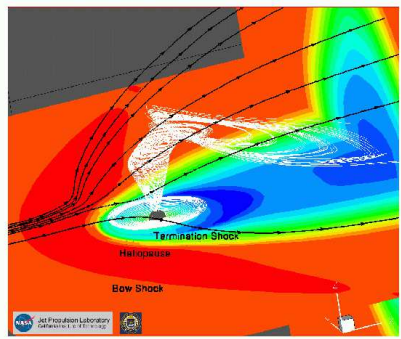}}
 \caption{Three-dimensional view of the global heliosphere. The color code shows the log of plasma density.
Black lines are the plasma
velocity streamlines. White lines follow the magnetic field lines.}
\label{fig1}
\end{figure}
\begin{figure*}[ht!]
\begin{minipage}[t] {0.5\linewidth}
\begin{center}
\includegraphics[angle=0,scale=0.3]{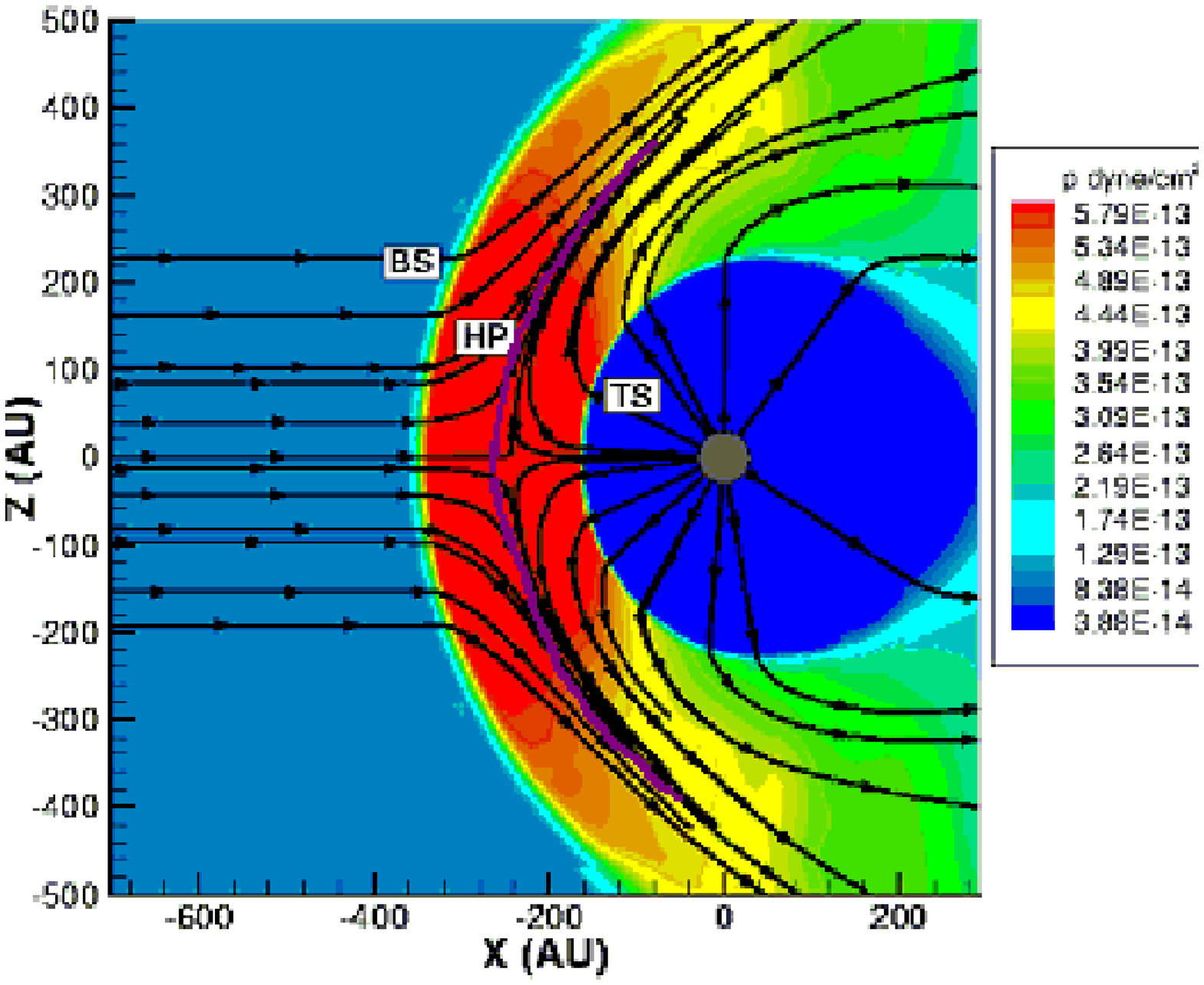}
\end{center}
\end{minipage} \hfill
\begin{minipage}[t] {0.5\linewidth}
\begin{center}
\includegraphics[angle=0,scale=0.26]{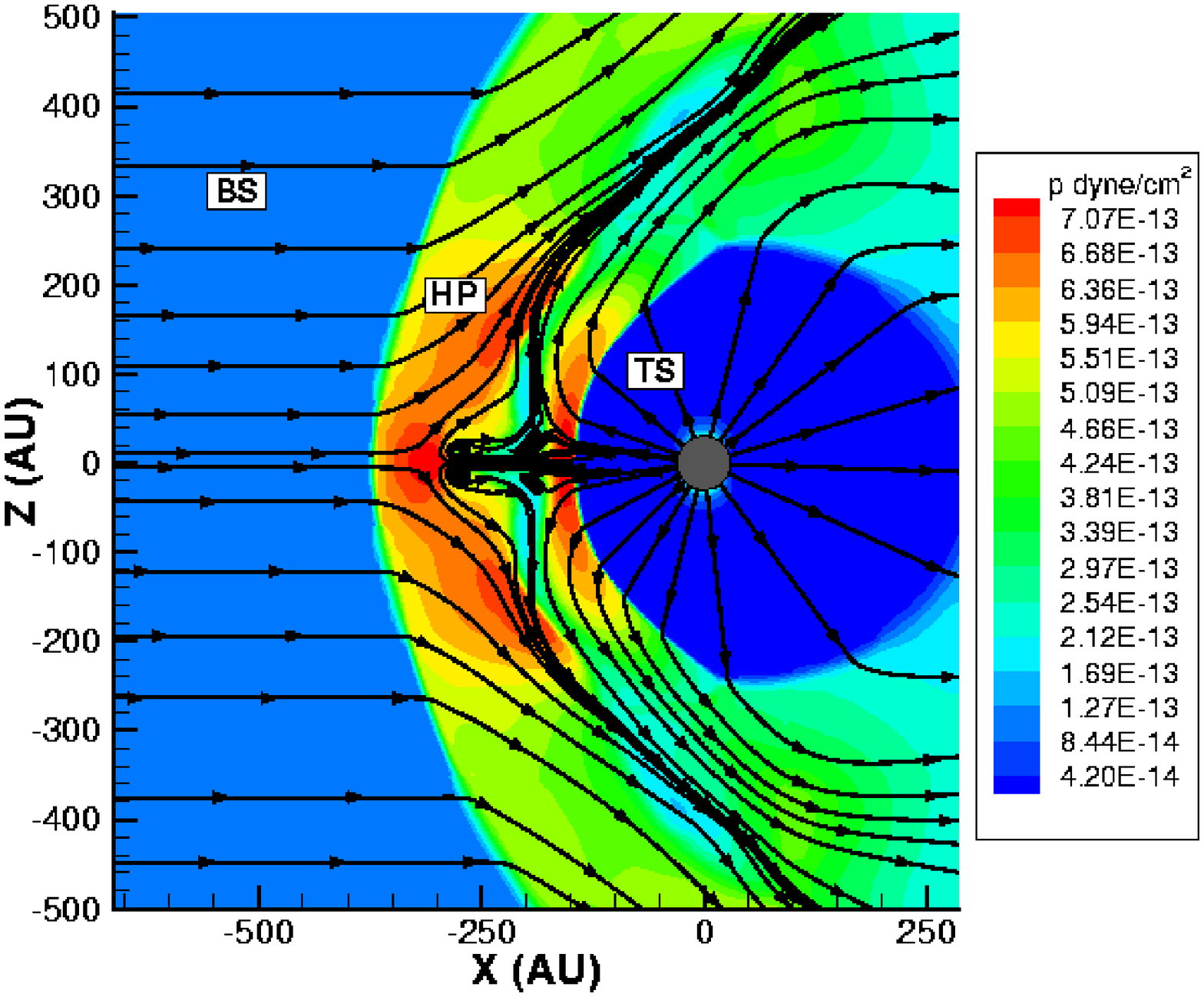}
\end{center}
\end{minipage} \hfill
\caption{Contours of pressure for (a) Case with no magnetic field (b) With magnetic field with
resolution of $0.75AU$ at the current sheet.}
\label{fig2}
\end{figure*}
\begin{figure*}[ht!]
\begin{minipage}[t] {0.5\linewidth}
\begin{center}
\includegraphics[angle=0,scale=0.3]{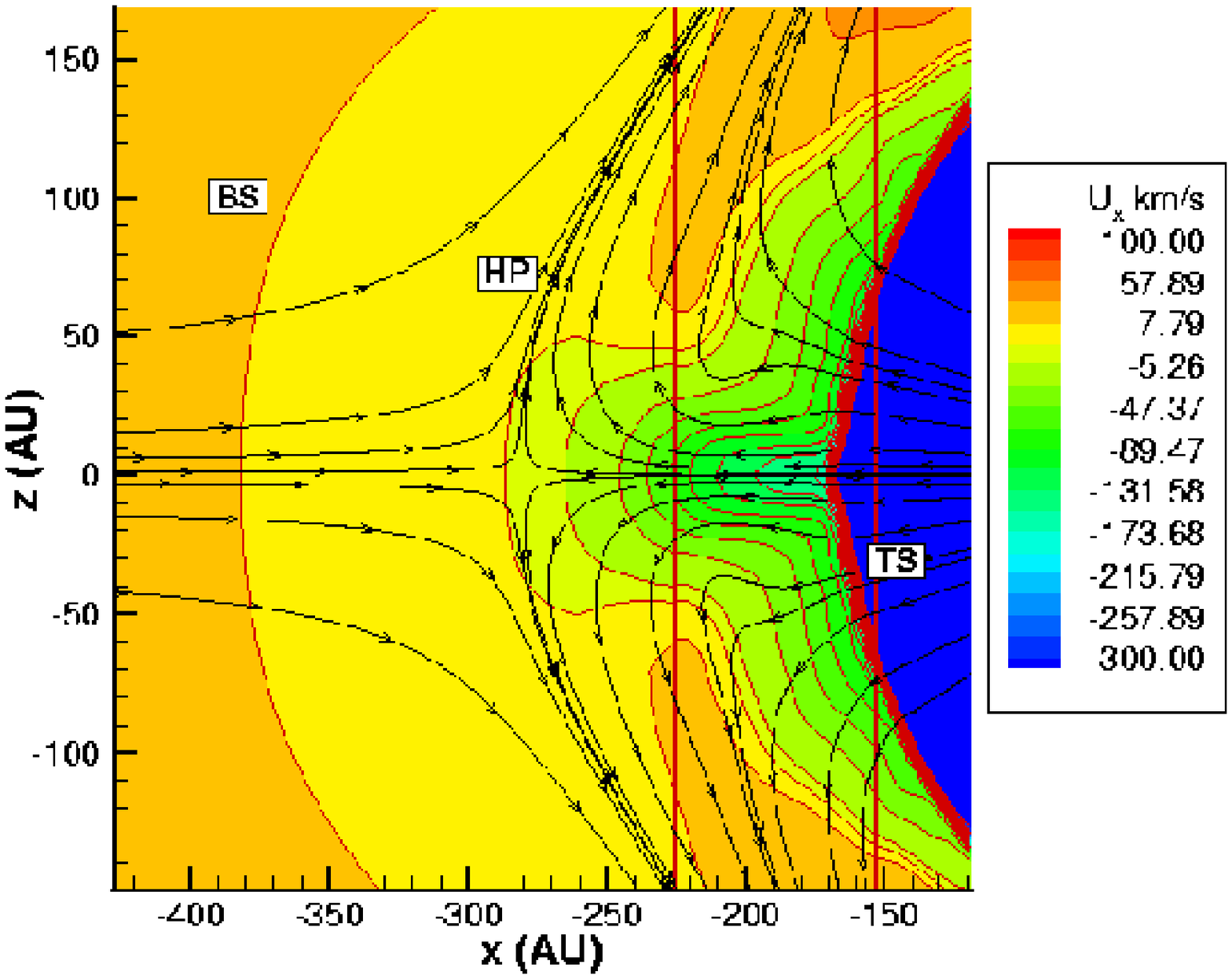}
\end{center}
\end{minipage} \hfill
\begin{minipage}[t] {0.5\linewidth}
\begin{center}
\includegraphics[angle=0,scale=0.3]{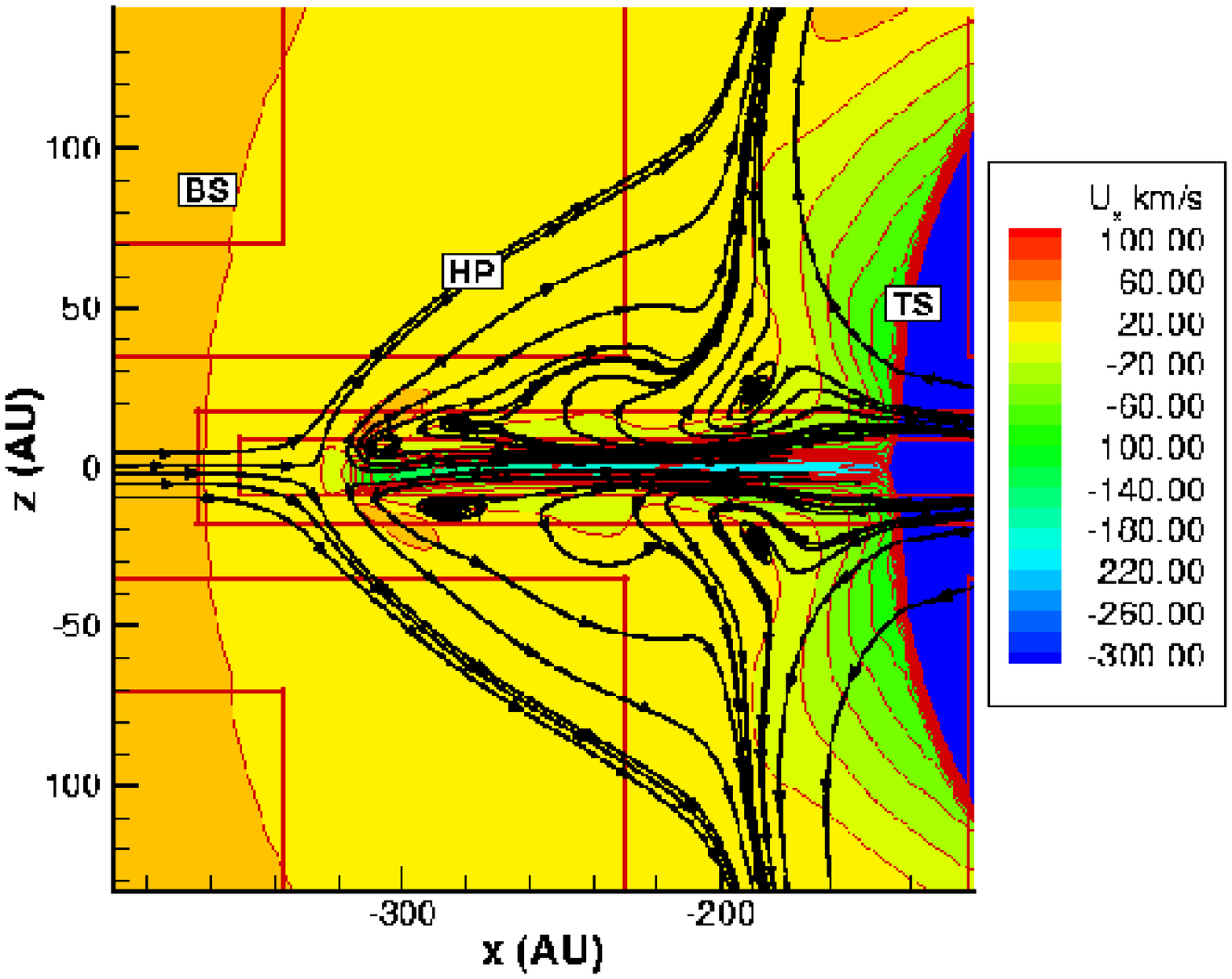}
\end{center}
\end{minipage} \hfill
\caption{Contours of velocity (a) Resolution of $3.0AU$; and (b) Resolution of $0.75AU$.}
\label{fig3}
\end{figure*}
It can be seen that in the case of lower spatial resolution, the jet at the current sheet is 
broadened. In that case, the current sheet remains in the equatorial plane as 
in previous studies with similar resolution\cite{linde,washimi}.

\section{Magnetohydrodynamic Instabilities}
\begin{figure*}[ht!]
\begin{minipage}[t] {0.5\linewidth}
\begin{center}
\includegraphics[angle=0,scale=0.3]{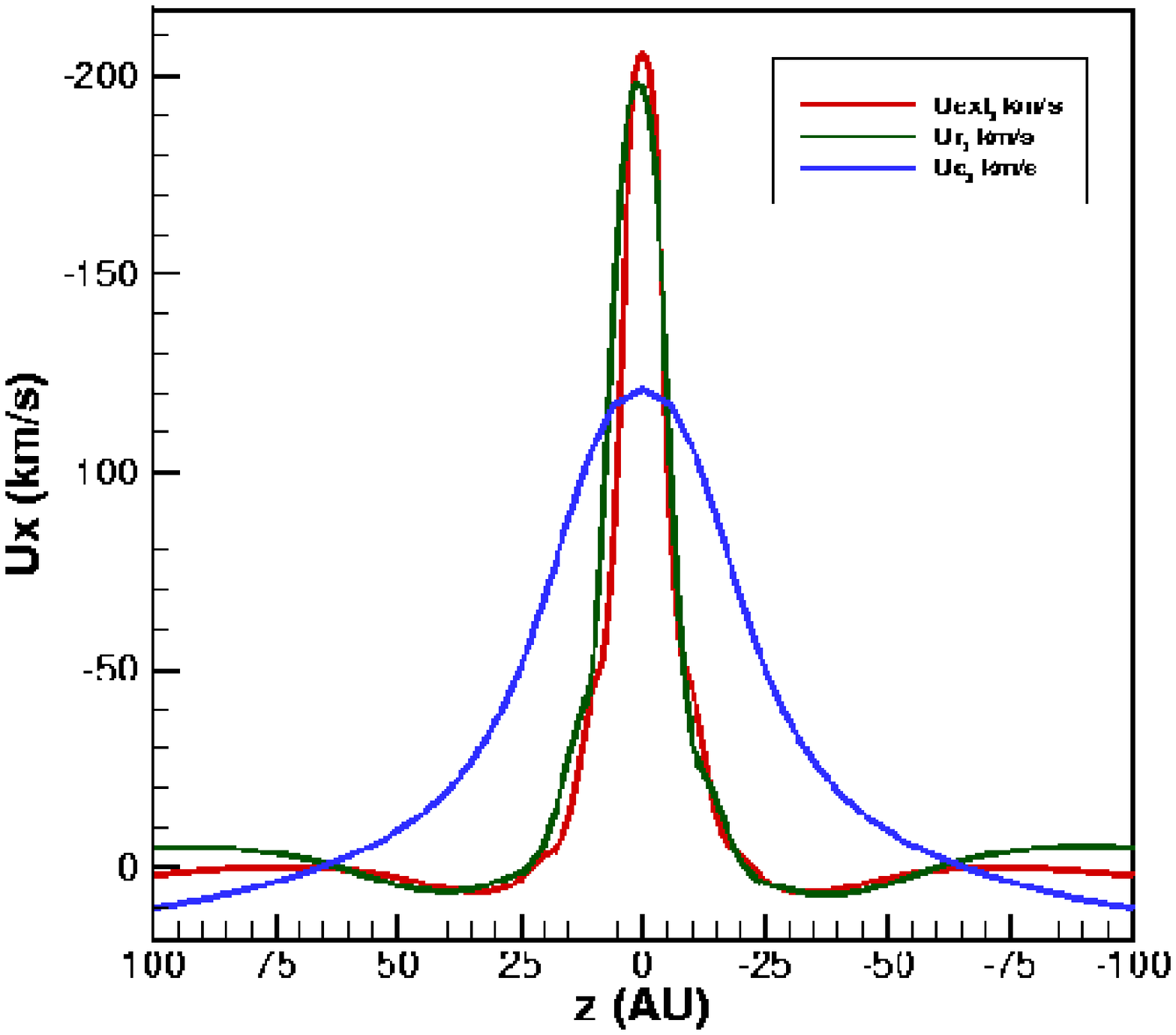}
\end{center}
\end{minipage} \hfill
\caption{Velocity profiles for the coarse ($3AU$), refined($1.5AU$) and super-refined-extended($0.75AU$) cases}
\label{fig4}
\end{figure*}
We found that at the TS there is a new effect at the current sheet: the converging flow near the equatorial plane 
creates a {\it de Laval nozzle}. A subsonic flow must accelerate where the 
streamlines converge and decelerate where flow lines diverge, while the opposite is true for a supersonic flow. The de 
Laval nozzle accelerates a jet at the current sheet for $150 AU$ beyond the TS.  At the TS the flow velocity decreases to 
Mach number $0.55$. Due to the acceleration of the flow past the TS, the Mach number increases to $1.1$. 
The velocity of the jet at the Laval nozzle is accelerated to $150 km/s$ and remains almost constant for the extension of the jet. 
Due to the difference of velocity between the jet and the surrounding flow, the current sheet becomes unstable 
due to a Kelvin-Helmholtz (KH) velocity shear instability \cite{opher,opher1}. At later times, the HP is highly distorted. 
To verify that this instability seen in the 3D MHD code is 
indeed caused by a KH type instability, we studied a much 
simpler configuration, analogous to that studied by Einaudi\cite{einaudi}. The code used was a 
2.5D compressible MHD code having a high spatial resolution of 0.07AU in the x direction. We 
verified (see also \cite{bettarini}) that in that case, as well as in ours, the jet develops a sinuous instability\cite{opher1}.
The growth rate for the instability as measured by Bettarini\cite{bettarini} is 
$\Gamma = 0.027 years^{-1}$ and the wavelength 
$\lambda = 25 AU$. These values are almost identical to what we observe in the 3D runs 
(for both cases with different spatial resolution), which reinforces the idea that the jet 
oscillation is due to the development of the KH instability. 

Figure 4 presents the velocity profiles for the three cases: the coarse (blue curve), the refined (green curve) and the 
super-refined-extended cases (red curve). The profiles for the velocity for the super-refined-extended 
and the refined case are almost identical. We can see that the width of the jet, in the super-refined-extended case, 
is independent of the grid resolution 
and is determined by the physical conditions, rather than by numerical resolution or numerical diffusion.

\section{Jet with Constant Neutral Background}
The neutral hydrogen component of the ISM interacts with the ionized component of the solar wind 
through charge exchange. Recently, we included the neutral hydrogen to our calculations as neutral background with constant 
density ($n_{H}=0.14 cm^{-3}$), velocity and temperature, coupled to the ionized component through charge exchange. 
Figure 5 shows the effect of the neutrals on the jet. The spatial resolution is comparable to the refined case\cite{opher}. 
Figure 5a shows the contours of velocity $U_{x}$. 
Figure 5b compares the line plot for the two cases with (red curve) and without (green curve) neutrals. 
We can see that with the inclusion of neutral atoms, the TS moves inward and there is a decrease in velocity 
by $100~km/s$ before the TS. At the TS, instead of the increase due to Laval nozzle right after the TS, 
the increase occurs further along the jet. 
Overall, the strength of the jet is reduced by the neutrals.We reserve a detailed discussion 
for a future paper.
\begin{figure*}[ht!]
\begin{minipage}[t] {0.5\linewidth}
\begin{center}
\includegraphics[angle=0,scale=0.32]{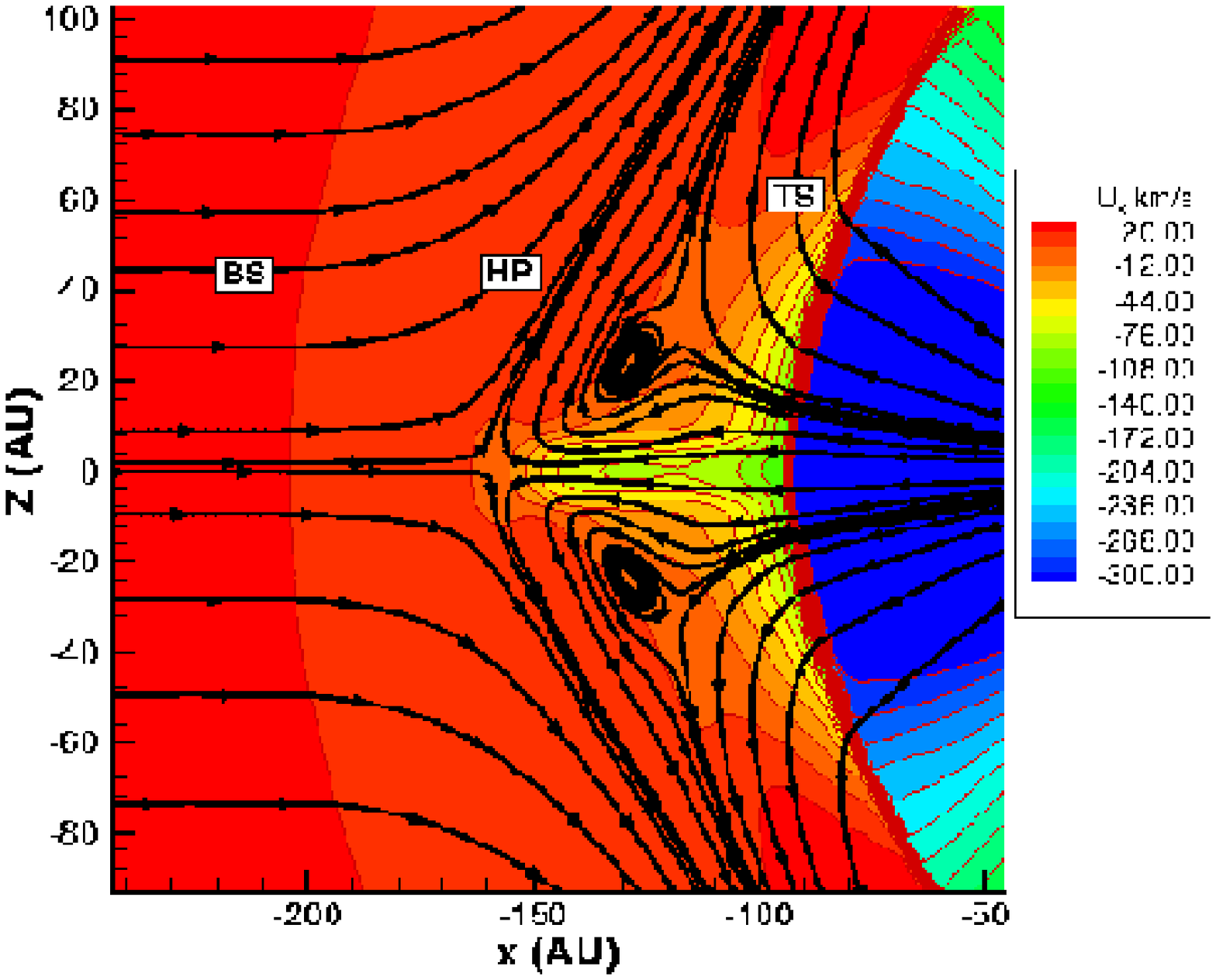}
\end{center}
\end{minipage} \hfill
\begin{minipage}[t] {0.5\linewidth}
\begin{center}
\includegraphics[angle=0,scale=0.3]{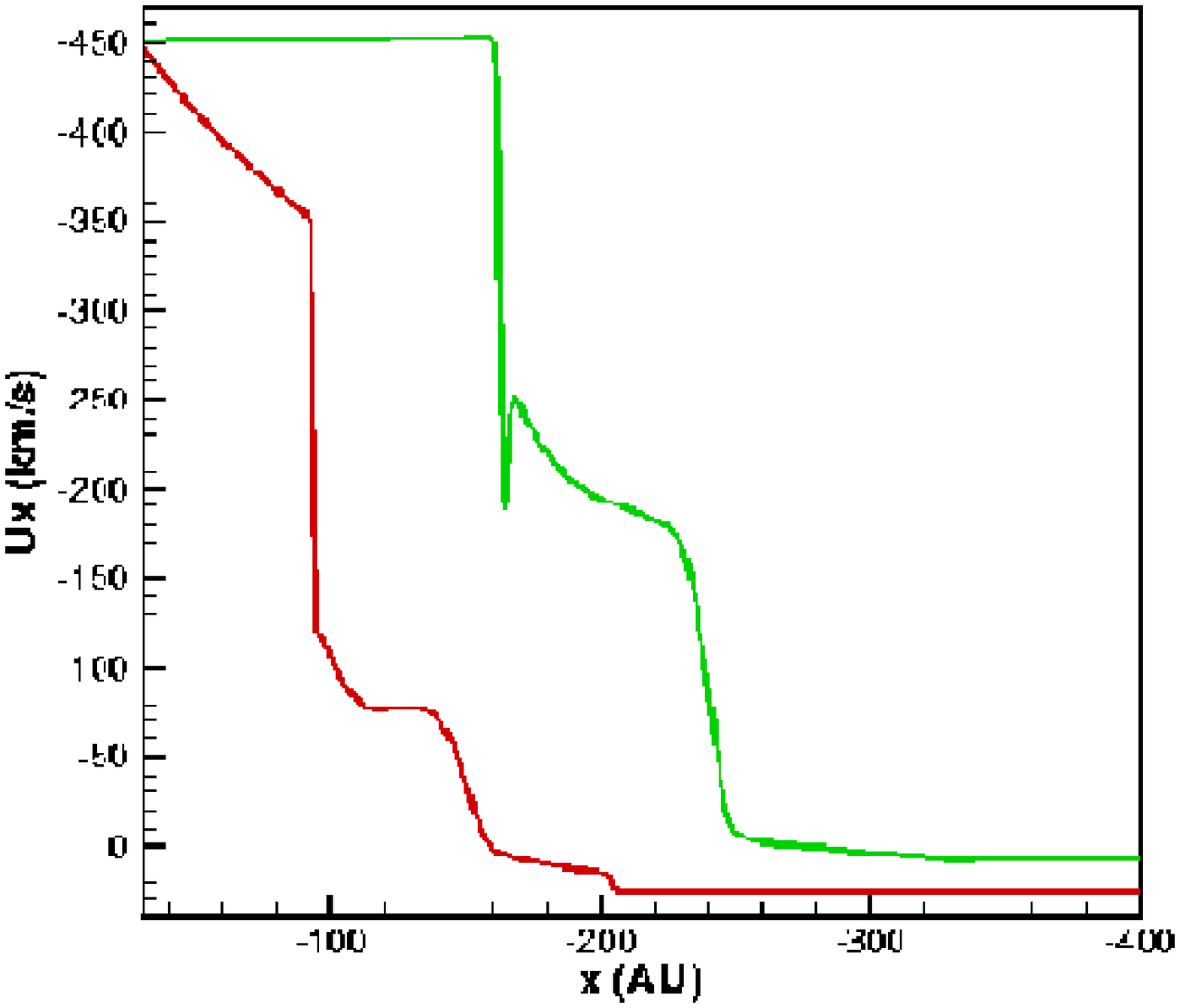}
\end{center}
\end{minipage} \hfill
\caption{(a) Contours of velocity $U_{x}$ at $t=24.9 years$. 
The black lines are the streamlines. (b) Line plot of the 
equatorial cut of $U_{x}$ vs. $x$. Two cases are shown: 
With no neutrals (red) and with neutrals (green)}
\label{fig5}
\end{figure*}

\section{Conclusions and Discussion}
We reviewed in this paper our recent results concerning the properties, the structure and dynamics of the Heliosheath. 
We discussed the presence of a jet of high speed flow at the current sheet that is only resolved with high spatial 
resolution. As we increased the spatial resolution, the profiles of velocity tend to a common shape. 
This result we believe, indicate that the width of the jet in the high resolution calculation, is independent of the grid 
resolution, and depends on physical conditions rather than on numerical resolution. We also showed that even under the 
presence of neutral atoms the jet is still present. The stability of the jet will be discussed in a future paper. 

High spatial resolution was a key factor for resolving the jet-sheet structure at the edge of the solar system. 
However, our model lacks important features such as the tilt of the magnetic axis with respect to 
the rotation axis, a self-consistent treatment of the neutral hydrogen fluid, and solar cycle 
effects were not included. The inclusion of a tilted heliospheric current sheet very likely will introduce qualitative 
changes in the picture 
represented in this paper. Nerney et al. \cite{nerney1} investigated 
analytically the solar cycle imprint on the Heliosheath. They predicted that 
magnetic polarity envelopes will be present in the Heliosheath with alternating polarities. 
The polarity of the magnetic envelopes 
reflects the polarity of a polar region of the Sun over the 11-year solar magnetic cycle. On a much finer 
scale, the magnetic field reverses polarity at least once per 25.5-day solar rotation in the strongly mixed 
polarity regions between the magnetic envelopes. 
Currently there is not a good enough model to be able to describe properly the Heliosheath.
From our recent results \cite{opher,opher1} we can state that the Heliosheath in nothing like we expected before. 
It exhibit high turbulence and back flows, high and slow velocities and gradients of density and pressure. More 
complete models are needed.

\begin{theacknowledgments}
This work is a result of the research performed at the Jet Propulsion 
Laboratory of the California Institute of Technology under a contract 
with NASA. The University of Michigan work was also supported by 
NASA. GT was partially supported by the Hungarian Science Foundation 
(OTKA, grant No. T037548). 
\end{theacknowledgments}

\end{document}